\documentclass[12pt,onecolumn]{article}
\usepackage{graphicx}
\usepackage{bm}
\usepackage{geometry}
\usepackage{cite}
\usepackage{times}
\usepackage{color}
\usepackage{a4wide}
\usepackage{amsmath}
\usepackage{csquotes}
\usepackage{subfigure}
\usepackage{hyperref}
\usepackage[nameinlink]{cleveref}
\usepackage{xcolor}
\usepackage{url}

\begin{document}

\title{\huge Provenance of Lyfe:\\Chemical Autonomous Agents Surviving through Associative Learning}

\author{Stuart Bartlett $^{1,2}$*, David Louapre $^{3,4}$\\
$^{1}$ Division of Geological and Planetary Sciences,\\
California Institute of Technology,\\
Pasadena, CA 91125, United States\\
Email: \href{mailto:sjbart@caltech.edu}{sjbart@caltech.edu}\\
$^{2}$ Earth-Life Science Institute,\\
Tokyo Institute of Technology, Tokyo 152-8550, Japan\\
$^{3}$ Ubisoft Entertainment\\
94160 Saint-Mand\'e, France\\
$^{4}$ Science \'Etonnante\\
75014 Paris, France\\
\href{https://scienceetonnante.com/}{https://scienceetonnante.com/}
}

\maketitle

\begin{abstract}
We present a benchmark study of autonomous, chemical agents exhibiting associative learning of an environmental feature. Associative learning systems have been widely studied in cognitive science and artificial intelligence, but are most commonly implemented in highly complex or carefully engineered systems such as animal brains, artificial neural networks, DNA computing systems and gene regulatory networks, among others. The ability to encode environmental correlations and use them to make simple predictions is a benchmark of biological resilience, and underpins a plethora of adaptive responses in the living hierarchy, spanning prey animal species anticipating the arrival of predators, to epigenetic systems in microorganisms learning environmental correlations. Given the ubiquitous and essential presence of learning behaviours in the biosphere, we aimed to explore whether simple, non-living dissipative structures could also exhibit associative learning. Inspired by previous modeling of associative learning in chemical networks, we simulated simple systems composed of long and short term memory chemical species that could encode the presence or absence of temporal correlations between two external species. The ability to learn this association was implemented in Gray-Scott reaction-diffusion spots, emergent chemical patterns that exhibit self-replication and homeostasis. With the novel ability of associative learning, we demonstrate that simple chemical patterns can exhibit a broad repertoire of life-like behaviour, paving the way for \textit{in vitro} studies of autonomous chemical learning systems, with potential relevance to artificial life, origins of life, and systems chemistry. The experimental realisation of these learning behaviours in protocell or coacervate systems could advance a novel research direction in astrobiology, since our system significantly reduces the lower bound on the required complexity for emergent learning.\\
\href{https://youtu.be/xZlXIVKBbcc}{Click here for video abstract}
\end{abstract}

\section{\label{sec:intro}Introduction and Conceptual Framework}
Ever since the outspoken chemist Ilya Prigogine introduced the concept, dissipative structures and their underlying phenomenology have beguiled, entranced and split opinions of researchers from a range of disciplines \cite{kond,prigogine}. The approximate definition of a dissipative structure is a coherent and discernible dynamical pattern that is maintained by external disequilibria over a finite time. The structure exists at a higher phenomenological or descriptive level to its microscopic constituents, and can thus be considered `emergent' \cite{anderson,goldstein,hoel,holland,rasmussen}.\par
The most frequently cited examples of dissipative structures are fluid convection cells \cite{ahlers,bartlett2014natural,bartlett2016maximum,manneville,pesch}, hurricanes \cite{emanuel}, dynamic surfactant structures such as micelles, vesicles and droplets \cite{bachmann,hanczyc2011metabolism,hanczyc2,mayer,segre,zhu}, stars and galaxies \cite{nozakura}, black holes \cite{curir}, and biological organisms \cite{kondepudi2,lineweaver,virgo2011thermodynamics}. In the fields of Artificial Life and the Origins of Life (OoL), dissipative structures are a key focus area, serving as metaphors for simple forms of life and providing clues to the non-life to life transition \cite{cafferty2019robustness,egbert2019steering,froese2014motility,hanczyc2011metabolism,lancet2018systems}. Numerous experiments and simulations in this field have illustrated the emergence of subsets of life's properties. For example, various oil droplet systems have been shown to exhibit chemotaxis, in which the motion of the droplets is powered through the consumption of a fuel compound, and the droplets' motion follows gradients in the concentration of that fuel \cite{cejkova2014dynamics,cejkova2017chemotaxis,cejkova2017droplets,hanczyc2003experimental,hanczyc2,hanczyc2011metabolism,holler2019droplet}. Self-replication is a common phenomenon in reaction-diffusion systems (RDSs) \cite{bartlett2014life,bartlett2016precarious,bartlett2015emergence,froese2014motility,lee2,lee3,lesmes2003noise,pearson,virgo2011thermodynamics} and vesicle-based structures, also known as protocells \cite{chen2004emergence,hanczyc2004replicating,jia2019membraneless,jordan2019promotion,joyce2018protocells,mann2012systems,ono2005computational,rasmussen2016generating,shirt2015emergent,sole2007synthetic}. Non-living or artificial dissipative structures also exhibit ecological behaviour including competition \cite{adamala2013competition,bartlett2014life,bartlett2015emergence}, homeostasis \cite{engelhart2016simple} and symbiosis \cite{bartlett2016precarious,froese2}. Artificial cell studies commonly follow the doctrine of autopoiesis \cite{maturana2012autopoiesis}, which is closely related to the `chemoton' concept \cite{ganti1975organization} and the container-metabolism-program framework \cite{rasmussen2008roadmap}, in that they seek integrated cellular structures comprising a boundary or membrane, a metabolic system for converting precursor `food' molecules into the components of the cell, and an information system that is normally analogous to a genetic system \cite{joesaar2019dna,joyce2018protocells,kurihara2015recursive,li2014synthetic,mann2012systems,qiao2017predatory,sole2007synthetic}. While this approach is philosophically sound, the simultaneous synthesis of all the required chemical components for such systems has proven to be prohibitively difficult, especially in `prebiotically plausible' scenarios, normally requiring very high precursor concentrations \cite{aubrey2009role,chandru2016abiotic}. Additionally, the physical driving forces that would compel a system of molecular components to exhibit the requisite dynamics that we associate with life are poorly understood \cite{krishnamurthy2017giving,krishnamurthy2018life}.\par
While pursuing a parsimonious, `straight-shot' from prebiotic physics and chemistry to biology is a logical starting point in seeking the OoL, given the often bizarre twists and turns that life's evolution has taken \cite{carroll2001chance,carroll2020series,gould1982exaptation}, it is also possible that the earliest forms of life were not simplified versions of extant life, but rather different in composition and organisation \cite{chandru2020polyesters,chandru2020prebiotic}. A natural question thus arises: what are the conserved quantities that we expect extant life, early life, and even extraterrestrial life, to exhibit? While definitions of life abound \cite{Cleland2007,cornish2020contrasting,mariscal2018life,mix2015defending,ruiz2004universal,vitas2019towards,witzany2020what}, and much previous work was inspired by the autopoiesis and chemoton concepts, a definition has recently been introduced that combines traditional and modern ideas from thermodynamics, biology, information theory and cognitive science \cite{Bartlett_2020}. This `four pillared' definition suggests that a living system must exhibit all of the following properties: 1) Dissipation (the system must be exposed to one or more thermodynamic disequilibria or free energy sources), 2) Autocatalysis (the system must exhibit or have the capacity to exhibit exponential growth of a representative size metric under ideal conditions), 3) Homeostasis (the system must possess regulatory or negative feedback mechanisms that can mitigate external or internal perturbations), 4) Learning (the system must have the ability to sense, store, process and exploit information).\par
While various model protocell systems have primitive genetic components, and simpler artificial chemical systems exhibit self-replication, chemotaxis, and homeostasis, emergent learning in simple artificial systems has remained an elusive goal. Chemical computing itself is a large discipline that includes the incredible achievements of DNA computing \cite{bar2002protein,benenson2012biomolecular,elbaz2010dna,qian2011neural,shapiro2006bringing}, computing using the Belousov-Zhabotinsky reaction \cite{adamatzky2002experimental,duenas2020vitro,gorecki2015chemical,hohmann1998learning,holley2011logical,parrilla2020programmable,steinbock1996chemical,sun2017combinational,wang2016configurable}, and more general approaches that exploit the computational universality of chemical reaction networks \cite{banzhaf1996emergent,banzhaf2015artificial,blount2017feedforward,chen2014deterministic,duenas2019chemistry,gasteiger1993neural,hjelmfelt1991chemical,hjelmfelt1992chemical,hjelmfelt1993pattern,hjelmfelt1995implementation,magnasco1997chemical,muzika2013control,poole2017chemical,soloveichik2008computation}. Chemical computation has also been exploited for the control of agent-like entities such as simple robots \cite{dale2010evolution}, but such robots are not emergent and hence their analysis is primarily a heuristic tool for the engineering of swarm intelligence (as opposed to understanding the OoL).\par
Despite the great strides described above, the problem of the emergence of a self-organised entity that satisfies a minimal definition of life including a basic learning ability, is still very much open. Complex protocells that are scaled-down versions of extant life provide extensive but somewhat limited guidance to the OoL, since there remains a gap between the passive, blind worlds of physics and chemistry, and the active and goal-directed world of biology. In the present work, we explored a minimal complexity, emergent system that exhibits all the features of minimal life definitions, including the four pillars of `lyfe' \cite{Bartlett_2020}, autopoiesis \cite{maturana2012autopoiesis}, and the ability to perform associative learning.\par
This result reduces the lower bound on the required system complexity of an autonomous chemical agent capable of learning. Our approach is also expandable to larger sets of variables or features. Given the vast diversity of prebiotic reaction networks, it is almost guaranteed that the required reactions can be found in a natural system, in the context of the OoL. The key challenge is understanding the conditions in which learning becomes the most stable dynamical process in a system (the learning has to feed back positively on the stability of the patterns performing the learning, it has to be dynamically favoured).\par
Our approach is based upon the soliton-like, spatially organised instabilities that form in non-linear chemical systems, collectively known as reaction-diffusion structures (RDSTs) \cite{awazu2004relaxation,bartlett2014life,bartlett2016precarious,bartlett2015emergence,budroni2017dissipative,epstein,epstein2,froese2014motility,froese2,gagnon2015small,gagnon2018selection,gray,gray1988brusselator,halatek,kuznetsov,kyrychko,lee,lee2,lee3,mahara,nishiura2001spatio,pearson,purwins2005dissipative,stich2013parametric,virgo2011thermodynamics}. The systematic understanding of RDSTs traces all the way back to \cite{turing}, who introduced their founding principles alongside the foundations of morphogenesis, which became one of the pioneering triumphs of mathematical biology. The RDS used in the present work is commonly known as the Gray-Scott model (GSM) \cite{gray,gray1988brusselator,lee,lee2,lee3,pearson}, which is an elaboration of the Selkov model of glycolysis \cite{sel1968self}. Experimental realisations of this system include the ferrocyanide-iodate-sulfite reaction \cite{horvath2016mechanism,szalai2008patterns,szalai2008patternformation}, and the rich dynamics of heterogeneous catalysis on surfaces \cite{bertram2003pattern,beta2003controlling,ertl1991oscillatory,ertl2008reactions,li2001turing,sachs2001spatiotemporal}.\par
The GSM is a simple 2-dimensional RDS that has been shown to exhibit several emergent, life-like properties. It involves two chemical species, denoted A and B, interacting through a simple auto-catalytic reaction $A+2B\to 3B$. Along with an appropriate supply mechanism for A, and removal mechanism for B, this reaction produces a variety of patterns, including self-replicating spots, able to progressively expand into regions of space \cite{pearson}. In addition, previous work has shown that adding the thermal dimension to this system (original versions were isothermal) reveals even more layers of emergent phenomena including competition (between RDSTs and convection cells) \cite{bartlett2015emergence}, and thermal homeostasis through symbiotic thermal regulation \cite{bartlett2016precarious}.\par


\section{The Role of Learning in the Biosphere and its Origins}
Although the GSM exhibits life-like characteristics, it lacks an essential feature of life: learning. Broadly speaking, learning can be defined as the ability of a system to record information about its environment, and process that information to modulate its behaviour. It is believed that learning is an essential pillar of life, providing a sensitivity to the environment that can improve the survival probability of the living system (this includes fields such as cognitive science, 4E cognition (embodied, embedded, enactive, and extended), perception-action cycles, measurement-feedback protocols, immune cognition, etc.) \cite{baluvska2016having,ben2009learning,ben2014physics,delgado1997collective,erez2017communication,farnsworth2013living,gershman2021reconsidering,ginsburg2010evolution,harada2000evolution,hopfield1994physics,kirchhoff2017there,manicka2019cognitive,marzen2018optimized,mitchell2009adaptive,sorek2013stochasticity,tkavcik2016information,watson2016evolutionary}. Although learning is sometimes viewed as specific to organisms having a nervous system, a large and rapidly expanding list of non-neural learning mechanisms illustrate its biological ubiquity, and it is hypothesised that learning may have played a role in life's origins \cite{bartlett2019probing,froese2018horizontal,michel2013life,walker2013algorithmic,walker2014top}. Among the different learning mechanisms, associative learning is particularly relevant in this regard. It can be defined as the ability of a system to detect and record \textit{correlated} features about its environment \cite{hebb1949organization,hopfield1982neural,hopfield1994neurons}. Note that such correlations can be either positive (high levels of variable M are associated with high levels of variable N), or negative (high levels of M are associated with low values of N, and vice versa).\par
The selective advantage provided by such learning mechanisms can be illustrated by the case of predator-prey interactions. The simplest strategy for a prey to escape a predator would be to trigger a response at the moment of the actual predator attack. This purely reactive behaviour could be called a \textit{direct} strategy. However, since predators are often relatively fast and the motor abilities of prey relatively limited, there is a survival incentive for prey species to be able to anticipate and react in advance.\par
To do so, prey species use their sensory systems to detect warning signs of predators. For instance, gazelle detect stalking predators such as cheetahs by visual detection, fish can sense sudden variations in light or fluid flow, etc. Detecting a predator ahead of time can then trigger a rapid response in order to escape. Such a \textit{pre-emptive} strategy is clearly advantageous since it allows more time to escape, as opposed to waiting for the actual predator attack. However, since such an escape manoeuvre is energetically costly, prey species have an incentive to escape only when there is an actual threat, and avoid false positive detections.\par
The simplest way to achieve an escape response would be to react when a given sensory stimulus (for instance level of sound, or fluid flow disturbance) exceeds a certain threshold. However, that threshold will depend on the environment (e.g., background level of noise, water turbulence, etc.). As those environmental conditions may slowly vary because of natural cycles or prey displacement, the threshold for reaction should be continuously adjusted. For instance, for a fish, a change of visibility caused by depth, turbidity or seasonal cycles should have an impact on the decision to trigger a fast escape when a given visual stimulus is sensed. Achieving such a behaviour as the environment changes is an \textit{associative learning} task: prey species need to learn the extent to which a stimulus is associated with an actual threat in their current environmental conditions, and update their behaviour if environmental conditions happen to change.\par
In this framework of predator-prey interactions, we consider three possible strategies: a direct reaction, a pre-emptive reaction, and associative learning. Although the benefit of such an associative learning ability in the case of a predator-prey interaction is quite clear, let us illustrate how this could also play a role in the resilience and adaptability of a proto-living chemical system.\par
Consider a system that can be damaged by a chemical species $T$ (a toxin), which is delivered at regular intervals. If the system has the ability to produce an antidote $N$, that degrades the toxin, the simplest defense strategy would be the \textit{direct} mechanism, where antidote $N$ is directly produced when toxin $T$ is encountered, for instance through a catalytic reaction:
\begin{equation}
T \to T + N,
\end{equation}
note that throughout this work, it is assumed that mass conservation is satisfied through additional species that are omitted from the reaction expressions for simplicity (in this case, a precursor from which $N$ is synthesised). Although such a direct network would be able to tune its response to the strength of the attack, it does not involve any anticipation. If the toxin is delivered in a sudden way at high concentration, antidote production might be too slow to avoid significant damage to the system.\par
Now suppose that the delivery of toxin $T$ is always preceded by the delivery of another chemical compound $S$. The presence of $S$ could serve as a stimulus signaling the occurrence of $T$ ahead of time. A possible \textit{pre-emptive} strategy would then be to start producing antidote $N$ as soon as stimulus $S$ is detected, so the antidote concentration has time to build up in advance. This pre-emptive mechanism is equivalent to triggering a fast start escape whenever a predator is detected, rather than waiting for the actual predator strike. In our chemical system, such a pre-emptive mechanism can be realized by a reaction that produces antidote $N$ and is catalyzed by stimulus $S$:
\begin{equation}
S \to S + N.
\end{equation}
Now there is an obvious drawback with this strategy: since antidote production is controlled by the presence of stimulus, it is not sensitive to the actual strength of the attack (in our case, the concentration of toxin $T$). Suppose that the relative concentrations of $S$ and $T$ depend on slowly varying environmental conditions (like temperature, pH, or the concentration of another compound) such that a given stimulus concentration would not always correspond to the same amount of toxin. We could even imagine that under certain conditions, the association disappears so that stimulus remains but toxin occurrences cease. A simple pre-emptive network, relying on stimulus concentration alone, would produce antidote irrespective of the actual toxin level, and would even keep producing antidote after the association between $S$ and $T$ disappeared. This unnecessary production of antidote would be detrimental, as it would incur a metabolic free energy cost, or lead to damage if the antidote itself is a mild toxin. This type of spurious antidote production would be analogous to a prey animal repeatedly triggering fast start escapes even in the absence of a real threat. Although many prey species of mammal do exhibit this type of nervousness (even humans can be easily startled), they are clearly not fleeing the scene at every hint of a predator's presence.\par
To avoid this, a more efficient chemical network would have the ability to detect and record whether, in the prevailing environmental conditions, $S$ is actually followed by $T$, and to what extent (the strength of the association). Such a network would then adapt if the environmental conditions were to change, for instance if the relative concentrations of $S$ and $T$ are evolving, or if the delivery of $T$ is suppressed while that of $S$ remains. Fulfilling such a requirement is an associative learning task: the network should first detect to what extent $S$ is followed by $T$, record that information and react accordingly, while being able to update itself whenever the link between $S$ and $T$ changes. We can hypothesise that for a large class of conditions, such an \textit{associative learning} network would confer greater resilience than any direct or pre-emptive network, providing a selective advantage for emergent systems capable of associative learning.\par
As described above, chemical computing is already a well-developed field, and approaches to chemical learning have been suggested \cite{hjelmfelt1991chemical,hjelmfelt1992chemical,hjelmfelt1993pattern,hjelmfelt1995implementation,poole2017chemical}. However, the approach used in the present work was inspired by the learning networks of \cite{mcgregor}. Using search methods inspired by evolutionary processes, these authors were able to find abstract chemical networks able to solve learning tasks (such as basic association, and the AB-BA task). However, their model used a very large parameter space (searched using an evolutionary algorithm), and it was somewhat difficult to interpret how to choose the values of the different parameters involved. Furthermore, the learning networks were not embodied within emergent chemical structures.\par
The present work aims to: 1) explore minimally complex reaction systems that clearly exhibit associative learning, 2) investigate the resilience of such networks compared to simpler strategies, such as the direct or pre-emptive networks described above, 3) assess whether such networks can be embedded within emergent, spatially distinct, self-organised chemical structures. Such phenomena would serve as an important guide to the conditions in which the basic properties of life can emerge. Crucially, this emergence occurs in the absence of complex molecules, peptides, nucleic acids, biochemical reactions, metabolic cycles or membrane structures, and hence lowers the bound on the necessary conditions for chemical, autonomous, cognitive agents.\par
In the following section, we present minimal chemical networks that efficiently solve the learning task described above, when placed in a well-stirred (0-dimensional) environment. We display plausible external conditions in which those networks outperform direct or pre-emptive strategies. In the second part of the paper, we extend those results to the 2-dimensional case, where the learning network is coupled to a self-replicating spot system from the GSM. This provides a realisation of a dissipative structure able to replicate and adapt to its environment using associative learning, a feat which, as far as we are aware, has not been achieved before.


\section{Chemical Learning Networks in Well-stirred, 0-dimensional Systems}
\subsection{Environment and learning task}
Let us first describe the task to be solved. Consider an environment where a stimulus $S$ and a toxin $T$ are regularly delivered with period $\mathcal{T}$, and a temporal separation $\Delta\mathcal{T}$ between them, such that $S$ can be used as a signaling cue for the later occurrence of $T$ (we choose $\mathcal{T}=1000$ and $\Delta\mathcal{T}=400$). We consider that $S$ is delivered as boluses of unit concentration, while $T$ is delivered as boluses of concentration $\epsilon$. Finally, we assume both $S$ and $T$ degrade exponentially with a short characteristic time ($\tau_S = 0.2\mathcal{T}$), so that they quickly disappear after each bolus.\par
\begin{figure}[h]
    \centering
    \includegraphics[width=0.8\linewidth]{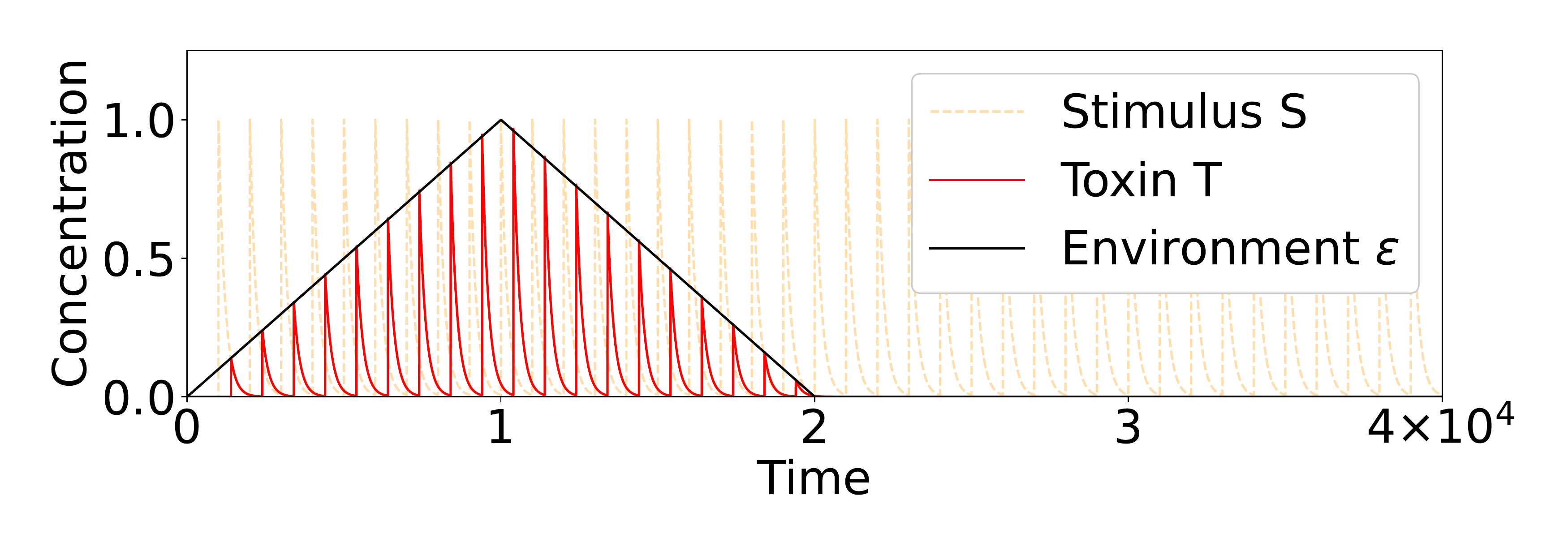}
    \caption{{\bf Dynamics of key compounds in learning task.} Stimulus and toxin bolus deliveries in the learning task to be solved. Delivery period is $\mathcal{T} = 1000$ and the time separation between stimulus and toxin occurrences is $\Delta\mathcal{T} = 400$. Stimulus boluses are always of unit concentration, while the magnitude of toxin boluses are controlled by the environmental parameter $\epsilon(t)$ (black line), which varies slowly over time. Both stimulus $S$ and toxin $T$ decay exponentially with a short characteristic time $\tau_S = 0.2\mathcal{T}$.}
    \label{fig:0D-environment}
\end{figure}

To model slowly varying environmental conditions that manifest as a modulation of the association between stimulus and toxin, the concentration $\epsilon$ of toxin boluses is allowed to slowly vary with time. We consider $\epsilon(t)$ linearly rising from 0 to 1 during the first 10 delivery periods, decreasing linearly during the next 10 periods, and staying at zero thereafter. This is illustrated in \autoref{fig:0D-environment}.\par
We now consider a chemical compound $B$, serving as a proxy for the presence of a proto-living chemical system. To achieve this, $B$ has an auto-catalytic dynamic, with an equilibrium concentration of unity, representing environmental capacity constraints. Evolution of the concentration of $B$, denoted $\psi_B$, is modelled by a logistic growth differential equation:
\begin{equation}
    \frac{d\psi_B}{dt} = \frac{1}{\tau_B} \psi_B (1 - \psi_B).
\label{eq:B_growth}
\end{equation}
Note that the logistic differential equation is a realisation of the following set of chemical equations: $B \longrightarrow B + B$ (autocatalysis) and $B + B \longrightarrow \emptyset$ (capacity constraint).\par
The growth rate of $\psi_B$ vanishes as it reaches its equilibrium value of 1, and $\tau_B$ represents the (minimum) characteristic growth time. We choose $\tau_B = 500$, so that the system has time to recover between each toxin delivery, but not completely.\par
To model the effect of toxin on our system, $T$ reacts with $B$, causing decay into inert waste products:
\begin{equation}
T + B \stackrel{k_{BT}}\longrightarrow \emptyset, \label{eq:BT}
\end{equation}
where the reaction constant $k_{BT}=0.5$. This moderate value ensures that the proto-living system is significantly damaged by the toxin at each delivery, without being completely destroyed by a single event.\par
\begin{figure}[h]
    \centering
    \includegraphics[width=0.8\linewidth]{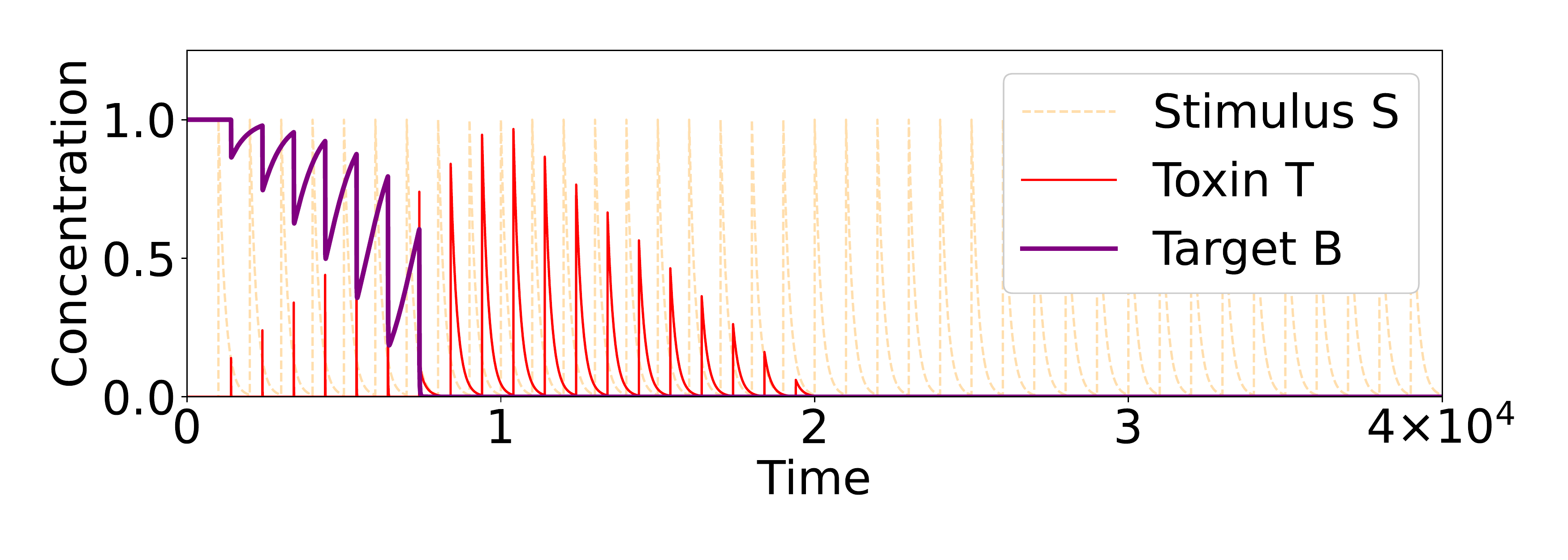}
    \caption{{\bf Dynamics of system in the absence of a defense mechanism.} Time evolution of the concentration of $B$ with toxin deliveries, in the absence of any defense network (network $\mathcal{N}_0$, see \autoref{tab:networks} for a summary of chemical equations used).}
    \label{fig:0D-no_network}
\end{figure}

The evolution of $B$ in the presence of stimulus and toxin delivery is illustrated in \autoref{fig:0D-no_network}. We see that the system gets progressively destroyed after each toxin bolus. Despite partial recovery between each delivery (due to the autocatalytic behaviour of $B$), the system (represented by the concentration of $B$) does not survive more than a couple of periods.\par
Suppose now that $T$ can be degraded by an antidote $N$, via a reaction into inert waste:
\begin{equation}
N + T \stackrel{k_{NT}}\longrightarrow \emptyset,	\label{eq:NT}
\end{equation}
where $k_{NT}=1$ so that the antidote is effective at removing $T$. To allow for functional anticipation mechanisms, we assume that the antidote is a long-lasting compound, and hence decays with a relatively long characteristic time $\tau_N = 5000$.\par
To implement the assumption that spurious antidote production is detrimental, we add a reaction in which the antidote degrades $B$ as well,
\begin{equation}
N + B \stackrel{k_{NB}}\longrightarrow \emptyset,	 \label{eq:NB}
\end{equation}
although with a much smaller rate constant $k_{NB}=0.0025$. There are two interpretations of this effect: 1) the antidote itself acts as a mild toxin, or 2) production of the antidote incurs a metabolic (free energy) cost to the prebiotic system, hence it consumes a certain quantity of $B$ via the above reaction. This will enforce the fact that antidote cannot be freely and arbitrarily produced, and must be used in a parsimonious and timely manner.\par

\subsection{Reaction network design}
Several defense networks can be considered for the production of antidote in response to the toxin. Below we describe a \textit{direct} network $\mathcal{N}_D$, a \textit{pre-emptive} network $\mathcal{N}_P$, and an \textit{associative learning} network $\mathcal{N}_A$

\underline{Direct network:}
The simple, `instantaneous' response of network $\mathcal{N}_D$ only involves one reaction in which the production of antidote $N$ is directly catalysed by $B$ in the presence of toxin $T$ (the relevant supply compound for $N$ is omitted):
\begin{equation}
B + T \stackrel{k_D}\longrightarrow B + T + N,  \label{eq:Direct}
\end{equation}
where reaction constant $k_D$ controls the rate of the response.\par

\underline{Pre-emptive network:}
This network $\mathcal{N}_P$, resembles the direct network, but the production of antidote $N$ is instead catalysed by the presence of stimulus $S$:
\begin{equation}
B + S \stackrel{k_P}\longrightarrow B + S + N, \label{eq:Preemptive}
\end{equation}
where $k_P$ controls the rate of the response.\par

\underline{Associative network:}
This network $\mathcal{N}_A$, involves two additional species: $M$, acting as a short term memory, and $L$, acting as a long term memory. Production of $M$ is catalysed by the presence of stimulus $S$:
\begin{equation}
S \stackrel{k_M}\longrightarrow S + M. \label{eq:assoM}
\end{equation}
To serve as a short-term memory, its characteristic decay time should be of the order of the expected separation time between the boluses of $S$ and $T$, hence $\tau_{M} = 200$.\par
The long term memory $L$ will encode the association between stimulus and toxin. To do so, $L$ will be produced from $M$ in the presence of $T$ (effectively acting as an AND gate on the simultaneous presence of $M$ and $T$), hence recording the fact that a bolus of $S$ has actually been followed by the occurrence of $T$:
\begin{equation}
M + T \stackrel{k_L}\longrightarrow L + T. \label{eq:assoL}
\end{equation}
Note that for the simple environment used here, it is possible to implement an associative network in which long term memory is stimulated only by the presence of T, precluding the need for short term memory M. This network is explored in the \hyperref[sec:supp_info]{Supplementary Information}, which also discusses why we retain short term memory in network $\mathcal{N}_A$.\par
To serve as a long term memory, the characteristic decay time of $L$ should be of the order of several multiples of the period $\mathcal{T}$, reflecting the rate at which the system will learn and adapt to a varying environment. Hence a value of $\tau_L = 4\mathcal{T} = 4000$ was chosen.\par
Finally, for this network production of antidote $N$ is catalysed by $B$ and the long term memory compound $L$, in the presence of stimulus $S$:
\begin{equation}
B + S + L \stackrel{k_A}\longrightarrow B + S + L + N. \label{eq:assoN}
\end{equation}
In this description of the associative network $\mathcal{N}_A$, three reaction rates $k_M$, $k_L$ and $k_A$ have been so far left unspecified, and can in principle be tuned to vary the dynamics and strength of the response. However, from the three reactions involved in the network, it can be seen that the amount of $N$ produced essentially depends only on the product $k_M k_L k_A$. Hence we choose to use only $k_A$ as a free parameter when performing sensitivity analyses. Without loss of generality, we set $k_M = 0.01$ and $k_L = 0.25$ (those values being chosen so that maximum concentrations of $M$ and $L$ remain of order unity).\par
The chemical equations used in each network are summarized in \autoref{tab:networks}.
\begin{table}[h]
    \centering
    \begin{tabular}{lll}
        \hline
        \textbf{Network} & & \textbf{Equation number} \\
        $\mathcal{N}_0$ & (No network) & (\ref{eq:B_growth})(\ref{eq:BT})(\ref{eq:NT})(\ref{eq:NB})  \\
        $\mathcal{N}_D$ & (Direct network) & (\ref{eq:B_growth})(\ref{eq:BT})(\ref{eq:NT})(\ref{eq:NB})+(\ref{eq:Direct})  \\
        $\mathcal{N}_P$ & (Preemptive network) & (\ref{eq:B_growth})(\ref{eq:BT})(\ref{eq:NT})(\ref{eq:NB})+(\ref{eq:Preemptive})  \\
        $\mathcal{N}_A$ & (Associative network) & (\ref{eq:B_growth})(\ref{eq:BT})(\ref{eq:NT})(\ref{eq:NB})+(\ref{eq:assoM})(\ref{eq:assoL})(\ref{eq:assoN})  \\ 
        \hline
    \end{tabular}
    \caption{List of chemical equations used in each network}
    \label{tab:networks}
\end{table}

\underline{Simulation:}
We simulate the dynamics of the networks and the environment with a set of coupled ordinary differential equations. If $\{C_i\}_{i=1..N}$ represents the set of $N$ chemical species, subject to $R$ reactions, the time evolution of concentration $\psi_{C_i}$ is given by
\begin{eqnarray}
	\label{eq:conc_ev_0D}
	\notag\frac{\partial\psi_{C_i}}{\partial t}= -\sum_{r=1}^{R}k_r\left[\alpha_{ri}'-\alpha_{ri}^*\right]\prod_{j=1}^{N}\psi_{C_j}^{\alpha_{rj}'},
\end{eqnarray}
where $\alpha_{ri}'$ is the left hand side stoichiometric coefficient for reaction $r$ and chemical species $i$ (the number of molecules of species $i$ entering as reactants into reaction $r$), and $\alpha_{ri}^*$ is the right hand side stoichiometric coefficient for reaction $r$ and chemical species $i$ (the number of molecules of species $i$ leaving as products from reaction $r$). Here we have ignored the role of temperature, and $k_r$ denotes the rate constant of reaction $r$. Reactions are simulated only in the forward direction, and possible reverse reactions are considered as extra, separate reactions from their forward counterparts. The exponential decay of each species is included as a reaction.\par
We simulated the networks coupled to the environment over 40000 time steps, starting with zero concentrations for all species except unit concentration for $B$. We use explicit time forward Euler integration ($dt=1$).\par

\underline{Optimisation:}
For each network, only one reaction rate is left unspecified, and can be tuned to vary the strength of the response: $k_D$ for the direct network $\mathcal{N}_D$, $k_P$ for the pre-emptive network $\mathcal{N}_P$, and $k_A$ for the associative network $\mathcal{N}_A$. Since our goal is to compare each of these networks at their maximum functionality, we searched for optimal values using line search optimisation, with the temporal averaged concentration of $B$ as the target to be maximised. Since $T$ degrades $B$, such an optimisation selects for values of reaction rates where the toxin causes minimal damage to the system. However, since $B$ is slightly damaged by excess levels of antidote $N$, an optimal solution will also avoid spurious production of $N$.


\subsection{Results}
\begin{figure}[h!]
    \centering
    \includegraphics[width=0.73\linewidth]{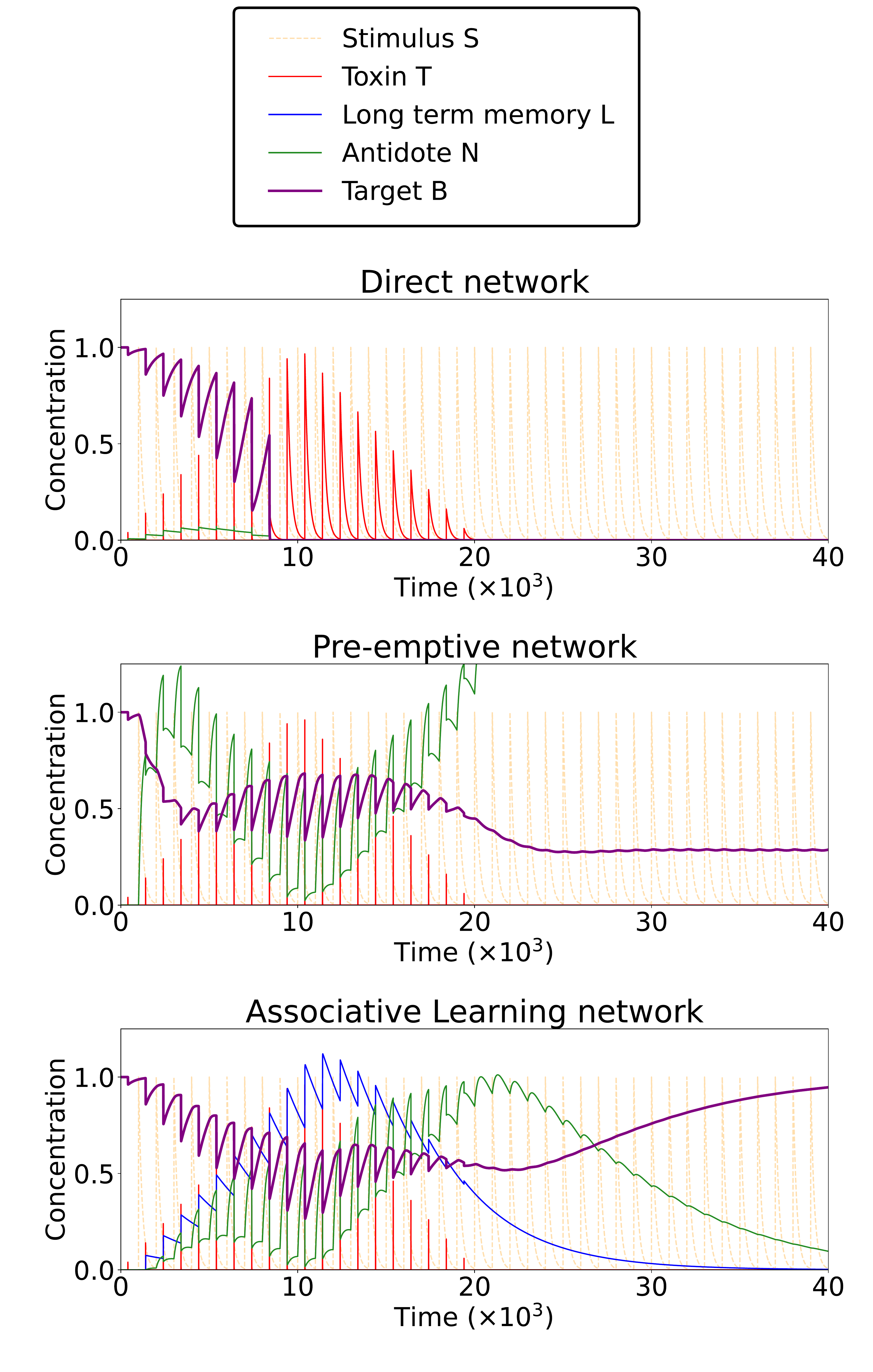}
    \caption{{\bf Concentration time series for the three networks.} Top: Direct network $\mathcal{N}_D$ with $k_D=0.1$. Middle: Pre-emptive network $\mathcal{N}_P$ with $k_P=0.005$. Bottom: Associative learning network $\mathcal{N}_A$ with $k_A=0.006$ (see \autoref{tab:networks} for a summary of chemical equations used in each network).}
    \label{fig:0D-3_networks}
\end{figure}

After appropriately tuning the reaction constants, we obtained the results displayed in \autoref{fig:0D-3_networks}, with $k_D=0.1$ for $\mathcal{N}_D$, $k_P=0.005$ for $\mathcal{N}_P$ and $k_A=0.006$ for $\mathcal{N}_A$.\par
We see that the direct network gets degraded after a few periods, and barely survives longer than when no network is present. The network is constrained by the production of antidote, and the toxin-induced decay of $B$ (which is required to catalyse antidote production). Loss of $B$ is thus too rapid to permit sufficient production of $N$. Note that there is an upper bound on the allowed rate constant $k_D$ (see discussion at the end of this section).\par
In contrast, the pre-emptive network starts to build up antidote concentration as soon as stimulus boluses occur. This leads to the accumulation of antidote, which is then at a sufficiently high concentration when toxin strikes, allowing an effective removal of $T$, and limited degradation of $B$. However, the amount of antidote produced is only linked to stimulus concentration and does not depend on toxin level. This leads to an excess of antidote when the toxin boluses are small, and to spurious antidote production in the second half of the sequence, when stimulus remains but toxin is no longer delivered. This creates a high residual concentration of antidote, that continuously damages the system, constraining the concentration of $B$ to a level significantly lower than its nominal equilibrium concentration $\psi_B=1$.\par
Finally, we see that the associative learning network is more resilient than both the direct and the pre-emptive networks. As the concentration of toxin boluses increases over time, long term memory $L$ accumulates and enables the anticipatory production of antidote $N$ when stimulus occurs. Furthermore, as the strength of the association between stimulus and toxin evolves, the long term memory concentration changes accordingly and promotes a response of appropriate magnitude. As the association disappears, long term memory decays to zero, as does antidote concentration, allowing a full recovery of the system towards its dynamic equilibrium. This provides an effective associative learning mechanism, recording information about the environment, using this information to react accordingly, while adapting behaviour when the external condition changes.\par
We can see that despite the learning mechanism, complete elimination of the toxin does not occur. This is because the system reaches a dynamic steady state through a feedback mechanism. Since the renewal of long term memory is catalysed by the presence of toxin, an excessive production of antidote would remove more of the toxin, and hence reduce the renewal of long term memory, which would subsequently decrease the antidote level. Conversely, if the initial antidote level is too low due to excessive toxin, $L$ would increase in response, with a concomitant increase in $N$. At steady state, the residual level of toxin provides just the necessary exposure to renew the long term memory.\par
\begin{figure}[h]
    \centering
    \includegraphics[width=0.8\linewidth]{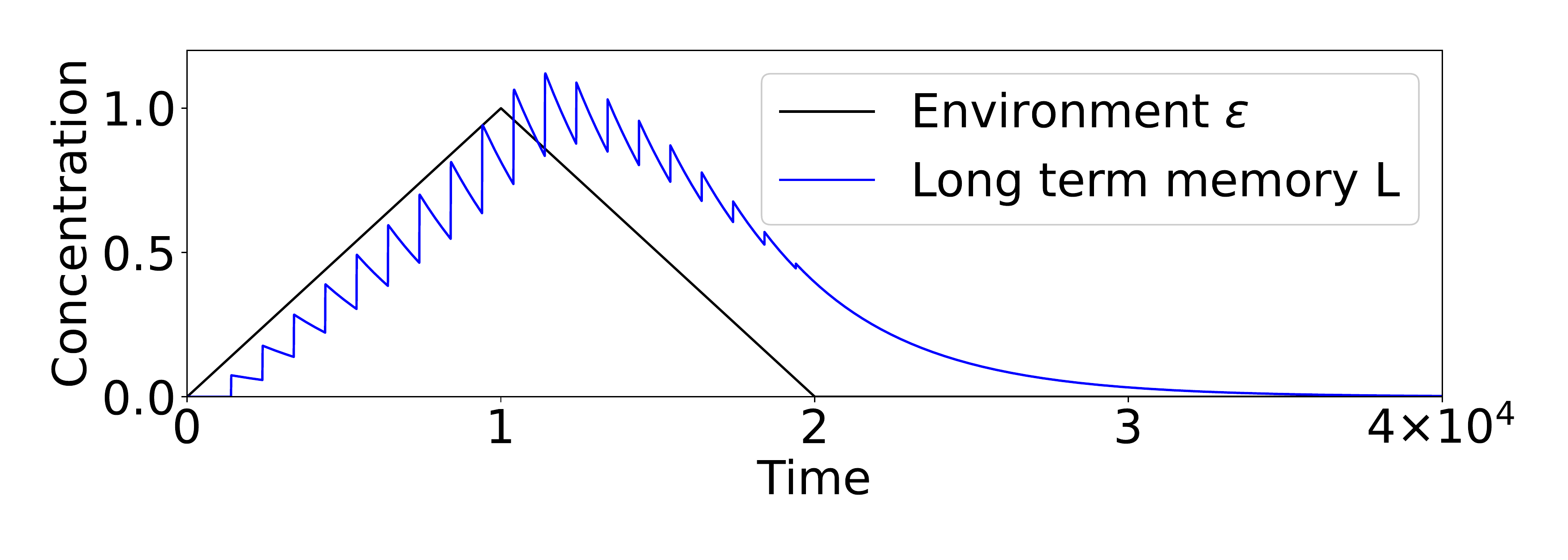}
    \caption{{\bf Dynamics of environmental parameter and long term memory.} Time evolution of the environmental parameter $\epsilon(t)$ controlling toxin bolus concentration, and long term memory $L$ for the optimal associative learning network.}
    \label{fig:0D-L_vs_epsilon}
\end{figure}

 
Moreover, variations in long term memory concentration closely mimic the variation of the environmental parameter $\epsilon(t)$ (that drives toxin bolus concentration), and displays a high temporal correlation with it, although with a slight lag, as can be seen in \autoref{fig:0D-L_vs_epsilon}. It shows that the associative network can effectively learn the value of $\epsilon(t)$ in a continuous fashion and adapt the system's behaviour, while periodically updating its knowledge about the environment.\par
\begin{figure}[h]
    \centering
    \includegraphics[width=0.9\linewidth]{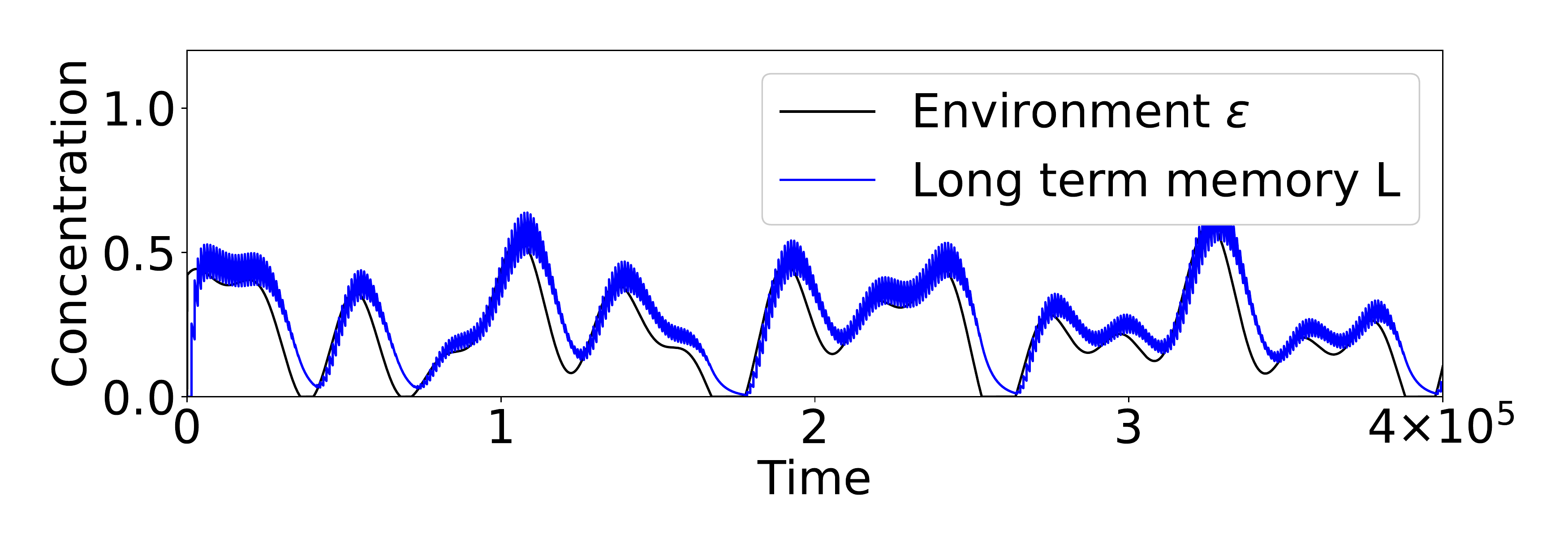}
    \caption{{\bf Long term memory concentration as a low pass filter.} Environmental parameter $\epsilon(t)$ varies with a more complex time evolution (computed as the sum of three sine waves of random amplitude, frequency and phase). The long term memory concentration acts like a low pass filter on $\epsilon(t)$. It approximates $\epsilon(t)$ with a slight lag related to the long term memory characteristic time scale, and also exhibits high frequency oscillations (appearing as the thick segment of the curve) due to the periodic bolus deliveries.}
    \label{fig:0D-L_vs_epsilon_complex}
\end{figure}

Since the network revises its memory state (the set of concentrations of network compounds could be considered the system's `composome' \cite{lancet2018systems,shenhav2004graded}) at every bolus (with period $\mathcal{T}$), and has a natural response time $\tau_L$, the evolution of $L$ over time can be seen as the result of a recursive, discrete, low-pass filter applied to the (time-varying) magnitude of toxin boluses $\epsilon(t)$. This is a natural interpretation since at each bolus, the network samples $\epsilon$ as the strength of the association between $S$ and $T$, and updates the level of $L$ accordingly. This is exactly the function of a discrete low-pass filter, since the discrete low-pass filter of a signal $x_n$ is a signal $y_n$ such that $y_{n+1} = (1-\alpha) y_n + \alpha x_n$. The parameter $\alpha$ governs the learning speed. For a low pass filter with characteristic time $RC=\tau$ and discrete signal sampled with time step $\mathcal{T}$, we find $\alpha = \mathcal{T}/(\mathcal{T} + \tau)$. This can be illustrated further with the case of an environment parameter $\epsilon(t)$ that varies with a more complex temporal pattern, as shown in \autoref{fig:0D-L_vs_epsilon_complex}.\par
\begin{figure}[h]
    \centering
    \includegraphics[width = 0.8\linewidth]{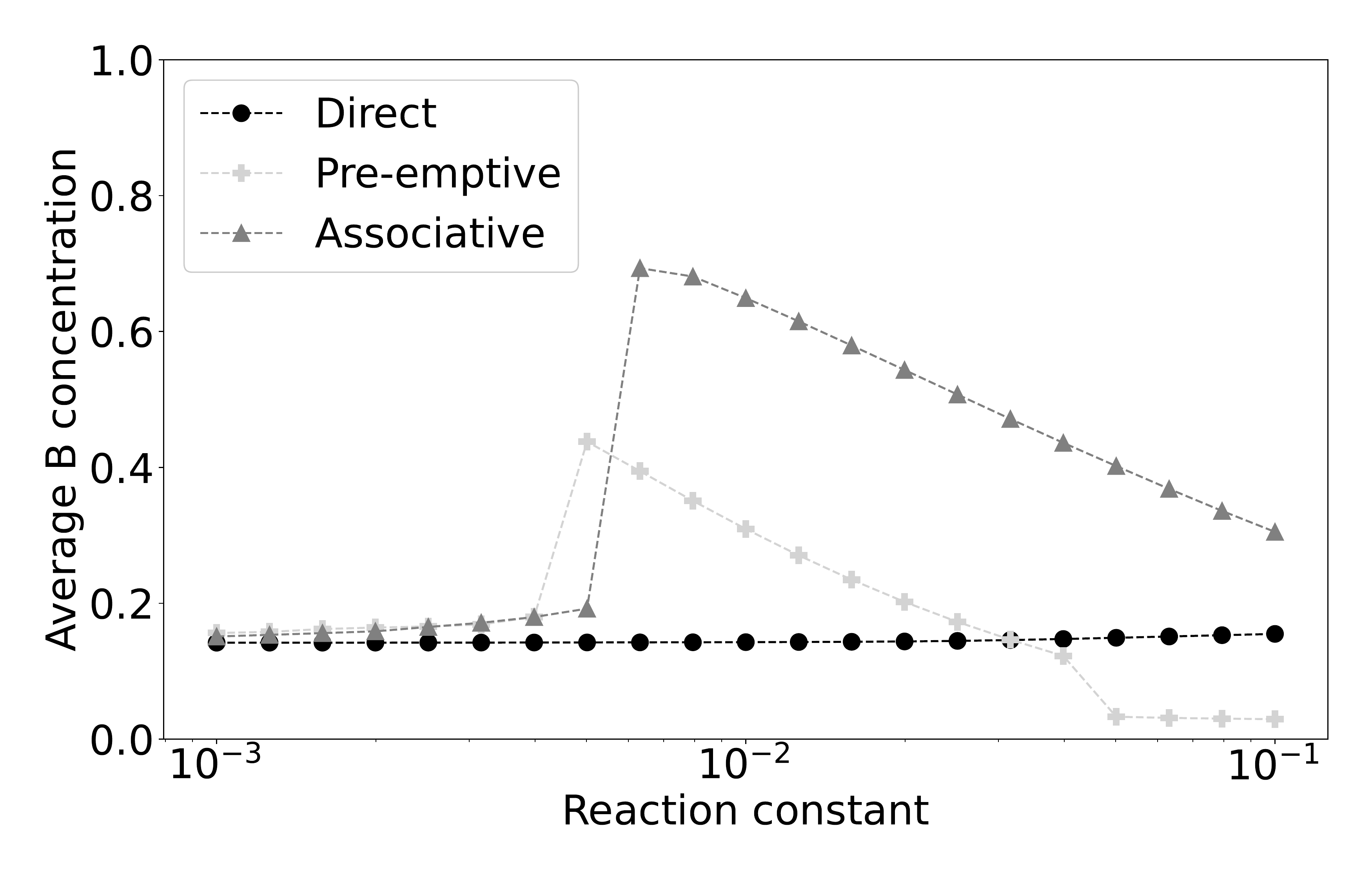}
    \caption{{\bf Performance variations as a function of reaction constant.} Performance of the different networks, measured as the temporal average of the concentration of $B$, as each key network reaction constant is varied. (See table \autoref{tab:networks} for a summary of chemical equations used in each network)}
    \label{fig:0D-optimisation}
\end{figure}

It is instructive to consider the performance of each network when the reaction constant is varied. This sensitivity analysis is summarised in \autoref{fig:0D-optimisation}. When the reaction constant is too low, all three networks perform poorly, since the system gets quickly destroyed by toxin. The direct network performs poorly for all values of the reaction constant $k_D$. When its reaction constant $k_P$ is sufficiently high, the pre-emptive network is able to neutralise the toxin. However, at higher values of $k_P$, its performance degrades because of damage caused by spurious antidote production at low or zero toxin level. Finally, the associative network is able to perform effectively for a larger range of reaction constant values, especially since it does not suffer from spurious production of antidote when the association between stimulus and toxin disappears. This illustrates that the associative network outperforms both the pre-emptive network and the direct network in a wide range of situations.\par
It can be seen that the performance of the direct network increases slightly with reaction constant, which is expected since a higher reaction constant allows that network to produce antidote at a higher rate when the toxin strikes. Increasing this reaction constant to even higher values would eventually allow that network to perform better. However this would require a very high production rate of the antidote, typically a hundred times greater than that of the pre-emptive and associative networks. Reaching such extreme production rates is unlikely in realistic settings, where such synthesis would be limited by various factors such as precursor availability or saturation effects. If arbitrarily high reaction constants were allowed, a direct network could eventually always perform as well as a pre-emptive or associative one. In the context of our predator-prey analogy, this would be equivalent to assuming that a prey animal can react and escape arbitrarily fast, in which case there is no need for it to learn to detect predators in advance. In practice, the speed at which a prey species can react is limited by environmental, physical and biochemical constraints, hence the reason that associative learning became a widespread and important trait in the biological world, providing a competitive advantage over simple reactive strategies.\par


\section{Chemical Learning Networks in 2D Reaction-diffusion Systems}
We now consider similar networks, but embed them in a 2D environment, coupled to a GSM. Hence there is a spatial dependence for the concentration of every species.
\subsection{Model description}
The model uses a $256\times128$ grid with a spacing of $dx=1$, and an isotropic diffusion coefficient for each species. The GSM consists of two chemical species A and B, subject to the chemical reaction $A + 2B \longrightarrow3B$, along with a decay mechanism for $B$ and a supply mechanism for $A$ \cite{gray,gray1988brusselator,lee,lee2,lee3,pearson}. The equations of motion for the concentrations of A and B are as follows:
\begin{eqnarray}
    \Dot{\psi_A} &=& D_A \nabla^2 \psi_A - \psi_A\psi_B^2 + f(1-\psi_A)\\
    \Dot{\psi_B} &=& D_B \nabla^2 \psi_B + \psi_A\psi_B^2 - (f+r) \psi_B,
\end{eqnarray}
with $D_A=0.2$, $D_B=0.1$, $f=0.03$ and $r=0.061$, a choice of constants known to give rise to self-replicating RDSTs, or spots \cite{pearson}. With this configuration, the spots are distinguished by local excess concentrations of $B$ (relative to the equilibrium concentration of 0), co-located with local deficits in the concentration of $A$ (relative to the equilibrium concentration of 1). We use the concentration of $B$ as an order parameter for visual representations.\par
To couple this GSM to the chemical networks described in the previous section, we use reactions that are identical to those of the 0D case, with the Gray-Scott $B$ component playing the role of the $B$ component in the 0D version. The only change we applied to the chemical reactions are for the associative network $\mathcal{N}_A$, where the production of short-term memory $M$ in the presence of $S$ now occurs only in the presence of $B$. This is to reflect the fact that short term memory can only be produced where spots are located:
\begin{equation}
B + S \stackrel{k_S}\longrightarrow B + S + M.
\end{equation}

\underline{Experimental relevance:}
In this numerical work, we used non-dimensional units (for time, space and concentration) to facilitate straightforward analysis and comparison with existing literature on numerical simulations of RDSs. However, we can easily translate these into meaningful experimental values. With a time scale of $dt = 1 \mathrm{s}$, and a spatial scale of $dx = 10^{-2} \mathrm{cm}$, our simulation would represent a 2D patch of approximately $2.56 \times 1.28 \mathrm{cm}^2$, with spots of approximately $1 \mathrm{mm}$ evolving over time scales of $\approx 10 \mathrm{min}$. With those values, the diffusion coefficients would be of order $10^{-5} cm^2/s$. Such values are in line with experimental work that has been carried out on RDSs using, for instance, iodate-ferrocyanide-sulfite reactions in a thin layer of polyacrylamide gel \cite{lee2}.\par

\underline{Bolus delivery:}
In this spatial system, the delivery mode of the boluses of $S$ and $T$ is a key consideration and two different cases were explored. In the so-called \textit{uniform} condition, boluses are delivered everywhere in a homogeneous fashion, as would occur for a 2D system sandwiched between porous plates, allowing the transverse diffusion of certain species (this is the way $A$ is supplied and $B$ removed in the classic GSM). Under this condition, all the spots come under equivalent attack from the toxin and are likewise homogeneously exposed to signals from the stimulus.\par
The second delivery mode is the \textit{boundary diffusion} condition, in which boluses are delivered on the boundary of the domain, and permeate via diffusion. For this condition, we use higher bolus concentrations, along with larger diffusion coefficients, to ensure that sufficient stimulus and toxin levels reach the center of the domain (see \autoref{tab:parameters}).\par
\begin{table}[h]
    \centering\scriptsize
    \begin{tabular}{lrl}
    Parameter & Value & Description  \\ \hline
    $C_S$ & 1 / 10 & Stimulus bolus concentration Uniform / Diffusion\\
    $D_S$ & 1 / 20 & Stimulus diffusion coefficient Uniform / Diffusion\\
    $D_T$ & 1 / 20 & Toxin diffusion coefficient Uniform / Diffusion\\
    $\tau_S$ & 100 & Stimulus decay time\\
    $\tau_T$ & 100 & Toxin decay time\\
    $D_N$ & 1 & Antidote diffusion coefficient\\
    $\tau_N$ & 400 & Antidote decay time\\
    $k_{TB}$ & 0.02 & Reaction constant T+B\\
    $k_{NT}$ & 1  & Reaction constant N+T\\
    $k_{NB}$ & 0.005 & Reaction constant N+B\\
    $k_M$ & 1  & Short term memory production reaction rate\\
    $k_L$ & 0.025  & Long term memory production reaction rate\\
    $D_M$ & 0.2  & Short term memory diffusion coefficient\\
    $\tau_M$ & 20  & Short term memory decay time\\
    $D_L$ & 0.2 & Long term memory diffusion coefficient\\
    $\tau_L$ & 400 & Long term memory decay time
    \end{tabular}
    \caption{Diffusion constants, characteristic decay times and reaction constants for the 2D networks. For stimulus and toxin diffusion coefficients, and stimulus bolus concentration, different values have been used for uniform and boundary diffusion conditions.}
    \label{tab:parameters}
\end{table}
For both conditions, we used a delivery period of $\mathcal{T}=100$ with an interval $\Delta\mathcal{T}=50$ between stimulus and toxin. The concentration of A is initialised homogeneously to $\psi_A^{t=0}=1$. To provide the initial seeds for the development of the spots, for each grid point $\psi_B^{t=0}=1$ with a probability of 0.35, and $\psi_B^{t=0}=0$ otherwise. To ensure the spots have time to fully develop before the toxin arrives, the boluses arrive only after 2000 time steps. Similar to the 0D case, the environment parameter $\epsilon(t)$, representing the magnitude of the toxin boluses relative to the stimulus boluses, rises linearly from 0 to 1 for 5000 time steps (from $t=2000$ to $t=7000$), then decreases linearly at the same rate (from $t=7000$ to $t=12000$) and remains at 0 thereafter.\par
The parameters of the reaction network were manually adjusted from the values found for the 0D case, and the three networks were compared at identical reaction rates ($0.15$ in the uniform case and $0.1$ in the boundary diffusion case). Simulations were performed using forward Euler integration (we used $dt = 0.1$ for the uniform case, and a smaller value $dt = 0.01$ for the diffusion case, because of the larger diffusion coefficient). To speed up the core calculations, the simulation was implemented on a GPU using HLSL compute shaders, and Unity3D software for interface and rendering.\par
To quantify the evolution of the spot system while being perturbed by $T$, we computed the spatial average of $\psi_B$ as a function of time. The number of spots was also counted using a thresholding and flood-fill procedure applied to the order parameter $\psi_B - \psi_A$ (threshold $=-0.3$).

\subsection{Results}
\label{sec:results}
\begin{figure}[h!]
    \centering
    \includegraphics[width=0.78\linewidth]{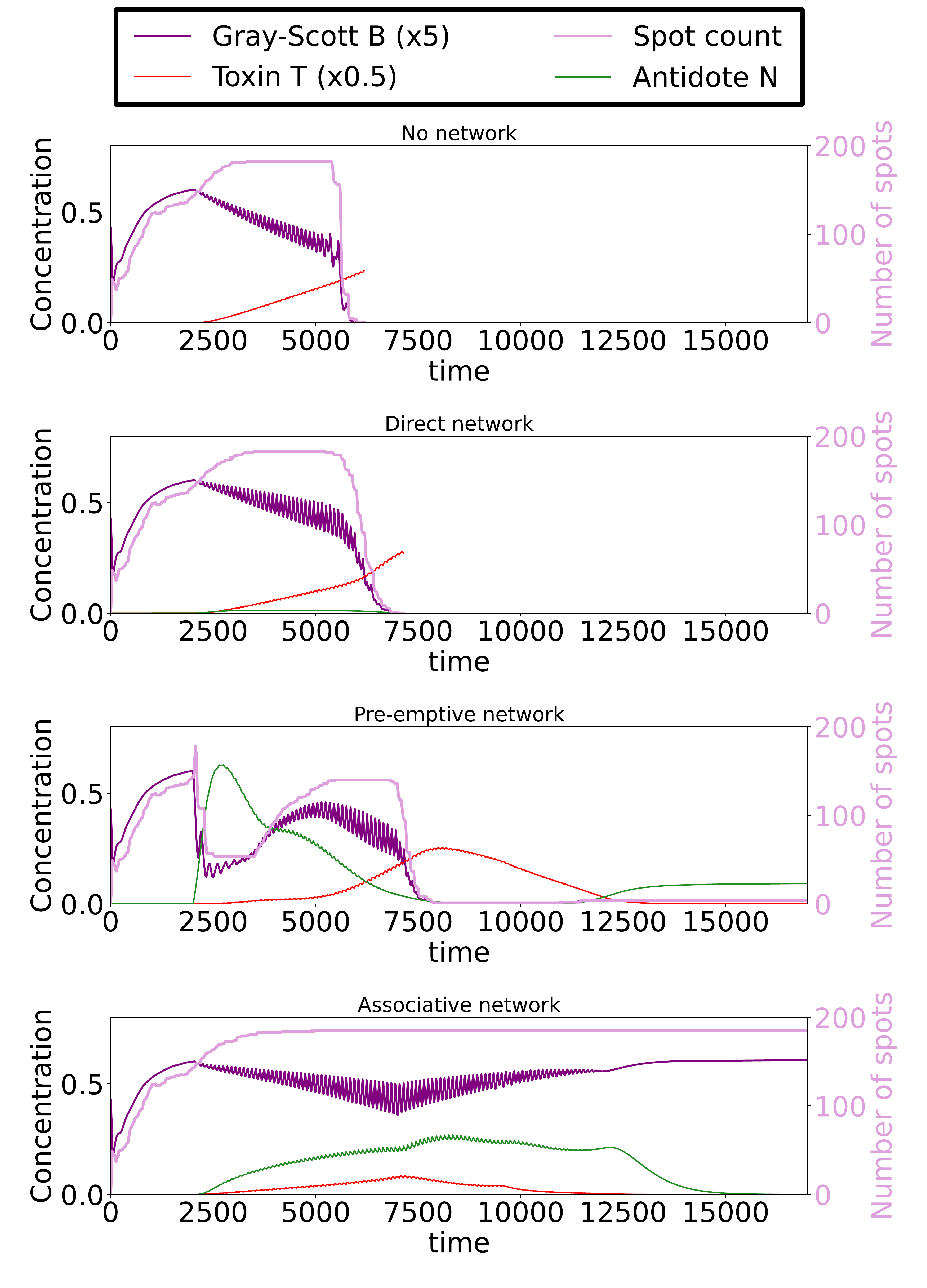}
    \caption{{\bf Dynamics of uniform delivery system.} Time series of the spatial average concentration of Gray Scott component $B$, Toxin $T$ and Antidote $N$, along with the number of RDSTs, in the case of uniform bolus delivery: 1) no network, 2) direct network 3) pre-emptive network, 4) associative learning network (time series of toxin and antidote concentration have been filtered to remove high frequency oscillations). See \autoref{tab:networks} for a summary of chemical equations used in each network.}
    \label{fig:uniform_results}
\end{figure}

\begin{figure}[h!]
    \centering
    \includegraphics[width=0.78\linewidth]{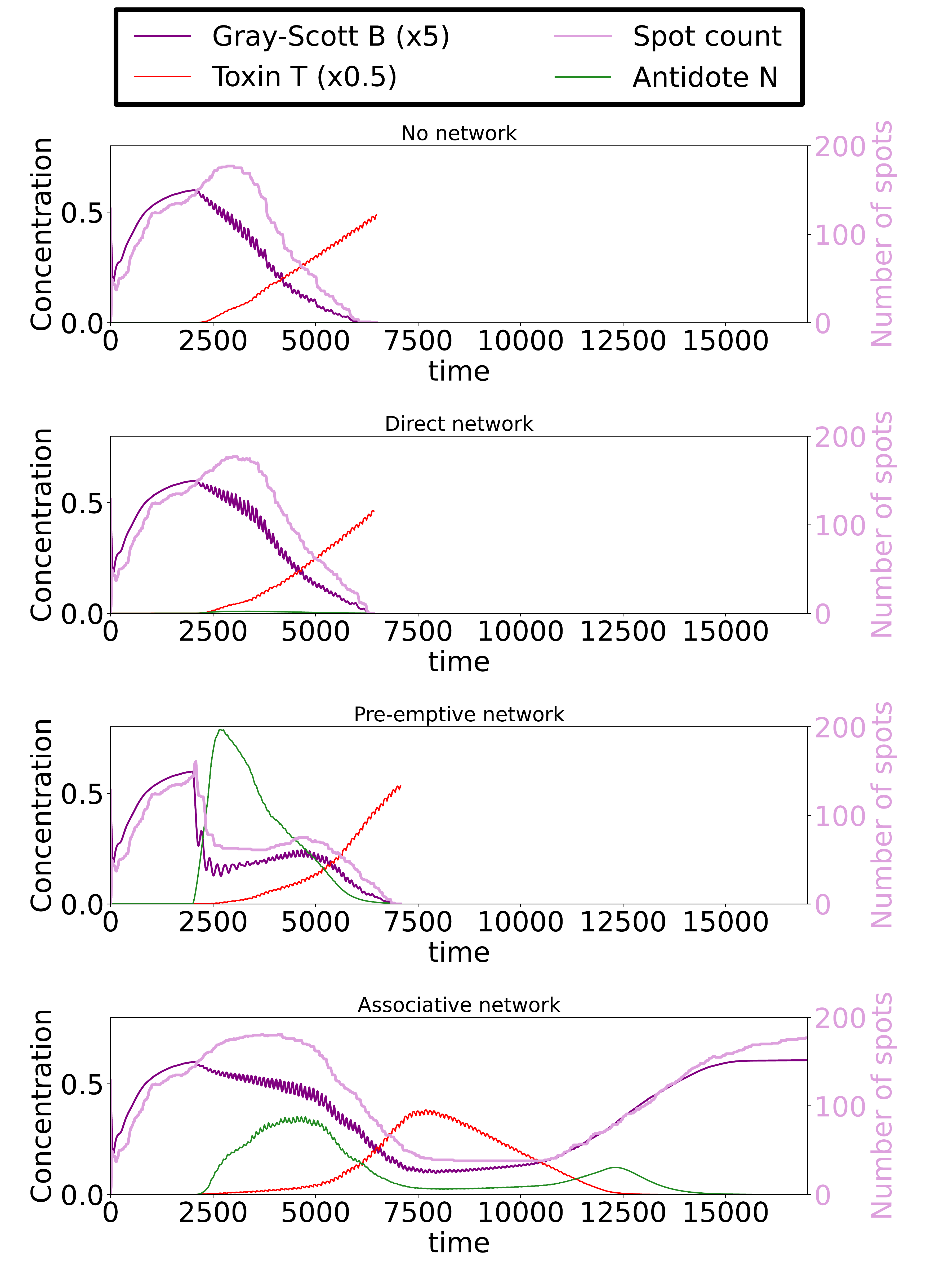}
    \caption{{\bf Dynamics of boundary delivery system.} Time series of the spatial average concentration of Gray Scott component $B$, Toxin $T$ and Antidote $N$, along with the number of RDSTs, in the case of boundary diffusion bolus delivery: 1) no network, 2) direct network 3) pre-emptive network, 4) associative learning network (time series of toxin and antidote concentration have been filtered to remove high frequency oscillations). See \autoref{tab:networks} for a summary of chemical equations used in each network.}
    \label{fig:diffusion_results}
\end{figure}

For both modes of bolus delivery (uniform and boundary diffusion), we simulated the three networks $\mathcal{N}_D$, $\mathcal{N}_P$ and $\mathcal{N}_A$ in addition to a control case with an empty network (no antidote production). Results for the uniform and boundary delivery modes are displayed in \autoref{fig:uniform_results} and \autoref{fig:diffusion_results}, respectively. Full video of all 8 simulations can be \href{https://www.youtube.com/watch?v=g1-SMI4Cl-A}{found here}.\par
For both delivery mechanisms, we can draw similar conclusions, which are analogous to what was observed in the 0D case. In the control case (without any antidote production), the spots get quickly destroyed by the toxin. Oscillations of the average concentration of $B$ can be observed, as toxin is periodically delivered and progressively damages the system, which only partially recovers in between boluses. This is analogous to the decreasing oscillations of $B$ that were observed in the 0D case (see \autoref{fig:0D-no_network}).\par
For the direct network $\mathcal{N}_D$, antidote is released when toxin boluses occur, but at an insufficient rate to prevent the complete destruction of the spots. As occurred in the 0D case, the system barely survives longer than when no antidote is produced.\par
For the pre-emptive network $\mathcal{N}_P$, antidote gets released when stimulus occurs, which constitutes an effective defense against the toxin. However, since it is only correlated with the stimulus level, the response is initially too strong when the toxin level is low, causing initial damage to the spots, and making them eventually disappear when the toxin level increases and causes additional damage, which is insufficiently mitigated by antidote production.\par
\begin{figure}[h!]
    \centering
    \includegraphics[width=0.78\linewidth]{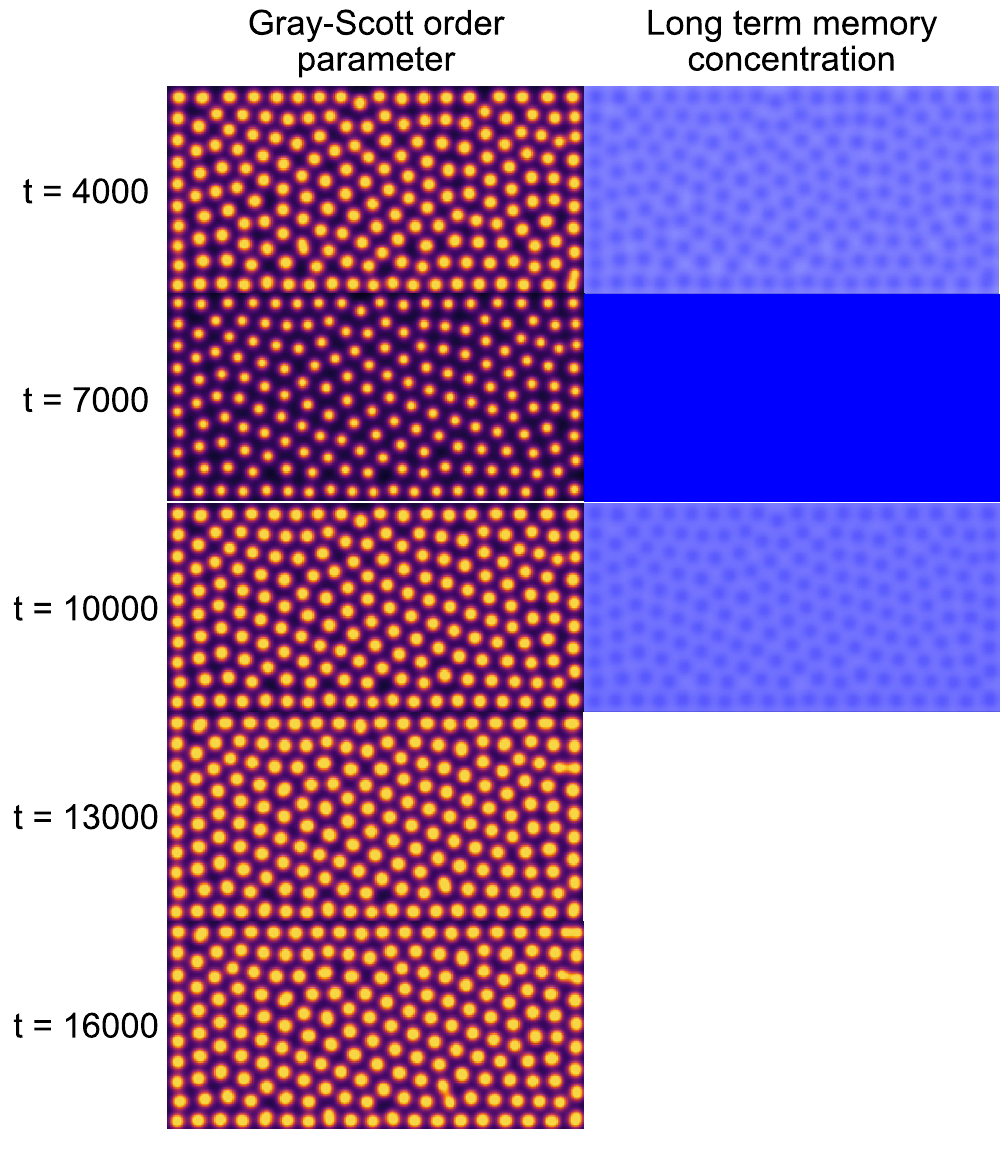}
    \caption{{\bf Spatial behaviour of uniform delivery system.} Snapshots of the Gray-Scott $B$ component and long term memory $L$ concentration at regular time intervals in the uniform delivery case, for the associative learning network $\mathcal{N}_A$. This sequence corresponds to the bottom panel of \autoref{fig:uniform_results}.}
    \label{fig:UniformAssociativeToxin}
\end{figure}
\begin{figure}[h!]
    \centering
    \includegraphics[width=0.78\linewidth]{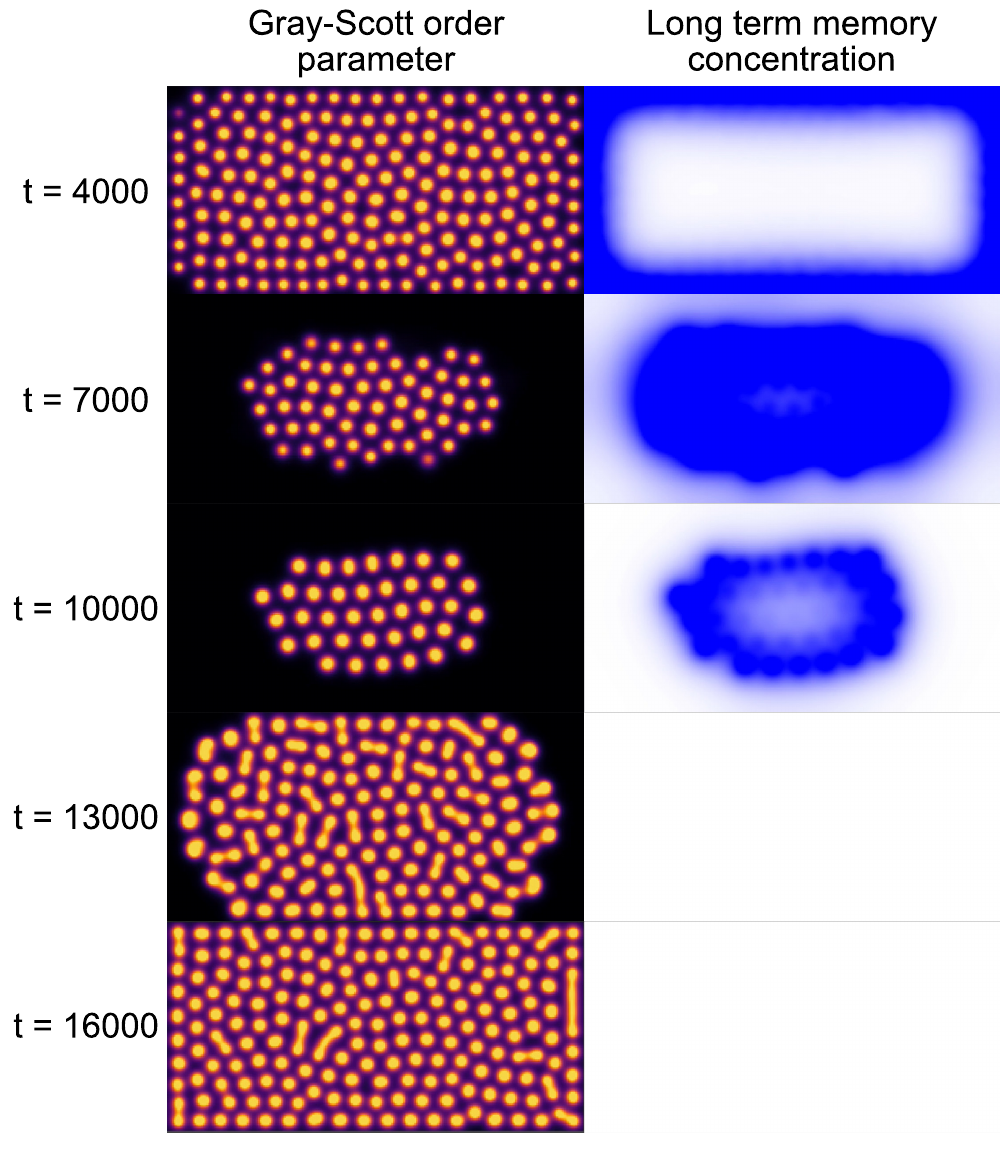}
    \caption{{\bf Spatial behaviour of boundary delivery system.} Snapshots of the Gray-Scott $B$ component and long term memory $L$ concentration at regular time intervals in the boundary diffusion delivery case, for the associative learning network $\mathcal{N}_A$. This sequence corresponds to the bottom panel of \autoref{fig:diffusion_results}.}
    \label{fig:DiffusionAssociativeToxin}
\end{figure}

Finally, for the associative learning network $\mathcal{N}_A$, production of antidote occurs only when and where it is necessary. In this condition, from the count of spots and the average concentration of $B$ (bottom left of \autoref{fig:uniform_results} and \autoref{fig:diffusion_results}), we can see that the spots get initially damaged by the toxin, until the progressive build-up of long term memory triggers the appropriate response and prevents further destruction. The spots then replicate autocatalytically and re-inhabit the space they previously lost (see \autoref{fig:DiffusionAssociativeToxin} and \autoref{fig:UniformAssociativeToxin} for snapshots of their spatial evolution).\par
Overall, the features observed in the 0D-case are re-capitulated in this spatial version. For the direct network, since antidote production is catalysed by the presence of toxin, the response will naturally be tuned to the magnitude of the attack. However the lack of an anticipation mechanism means that such a network would work only with a much higher reaction constant involving a very high instantaneous production rate of antidote. The pre-emptive network is able to produce antidote before toxin occurs but at a lower rate. This response is only proportionate to the level of stimulus, not the toxin. Hence it lacks the ability to tune itself to the actual magnitude of the toxin concentration. A sensitivity analysis revealed that no value of the reaction rate $k_P$ can lead to a survival of the system. A higher reaction constant is actually detrimental in the initial stages, when stimulus is high and toxin low, while a lower reaction constant leads to better survival at the beginning but increased damage when the toxin level becomes higher.\par
In contrast, we can see that the associative learning network combines both the anticipation mechanism of the pre-emptive network, and the self tuning ability of the direct network. A sensitivity analysis revealed that the network performs equally well for almost any value of the reaction rate $k_D$. As long as it is above a minimum of approximately 0.08 (for the uniform case), survival of the system occurs.\par
\begin{figure}[h]
    \centering
    \includegraphics[width=0.78\linewidth]{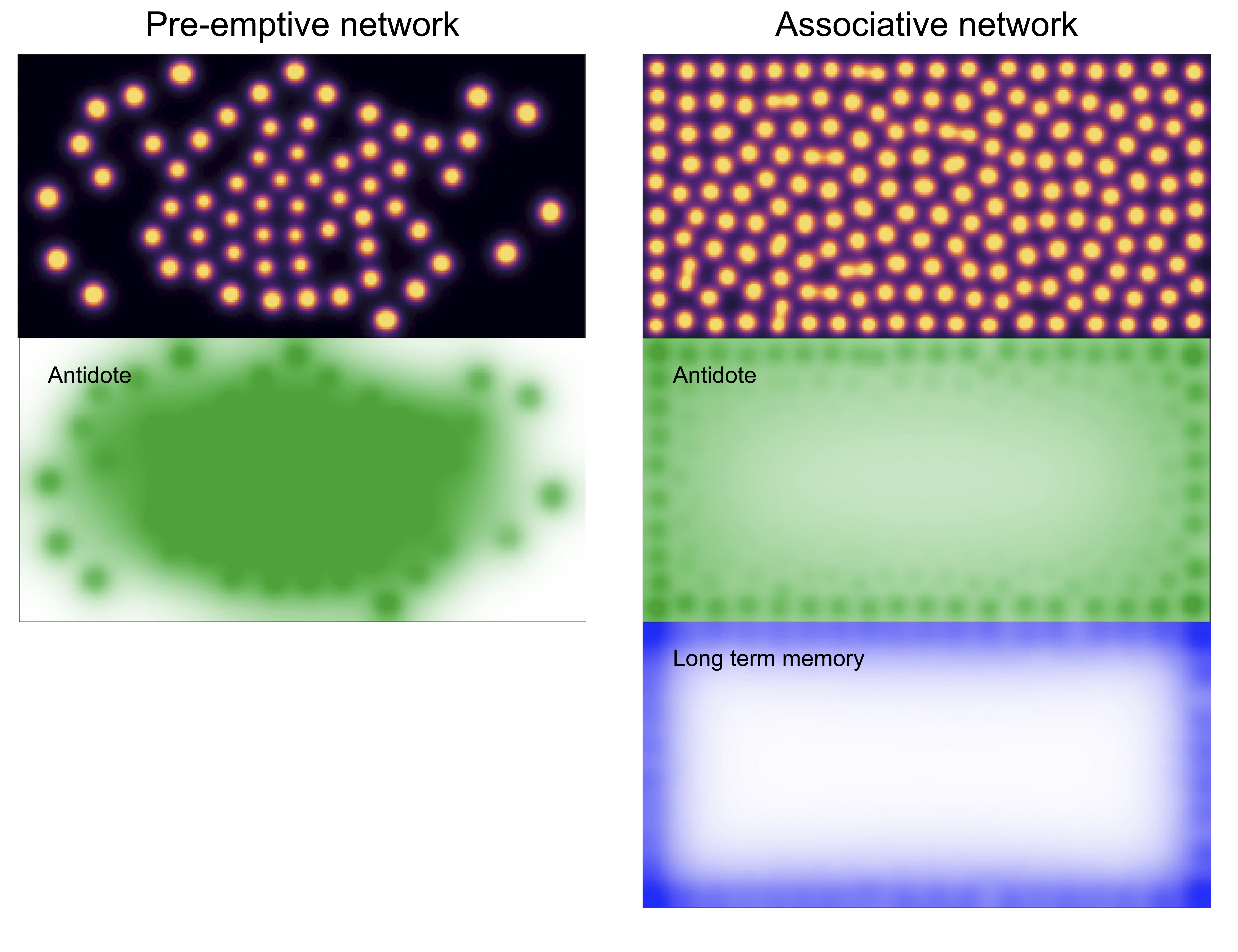}
    \caption{{\bf Comparison between pre-emptive and associative concentration fields.} Concentration map for pre-emptive and associative network in the boundary diffusion case, at the same time stamp ($t=3010$, after 10 toxin deliveries.) For the associative network (right), antidote production is located at the boundary, as dictated by long term memory concentration. For the pre-emptive network (left), antidote is produced by all spots in comparable amounts.}
    \label{fig:spatial_difference}
\end{figure}

To illustrate the self-tuning of the associative network, and how it impacts its spatial differentiation abilities, it is instructive to compare it with the pre-emptive network. \autoref{fig:spatial_difference} shows a concentration map after 10 toxin deliveries. We can see that for the associative network, antidote production is located on the boundary, as dictated by long term memory concentration. However for the pre-emptive network, antidote is produced by all the spots at similar rates, irrespective of their location. On the boundary, the antidote level is too low to resist high toxin levels. In the center, that same amount of antidote causes spots damage, since there is no toxin to degrade, and excess antidote is detrimental.

\section{Discussion}
The results of the previous sections demonstrate a rudimentary associative learning ability of simple catalytic reaction networks. This was achieved using a pair of memory compounds with different characteristic time scales allowing external correlations to be encoded. In addition, the synthesis of long term memory effectively proceeded through a logical AND operation on the presence of short term memory and the presence of toxin. Hence, simple catalysis, an appropriate separation of timescales, and associative synthesis reactions appear to be the necessary and sufficient conditions for chemical associative learning \cite{blount2017feedforward,mcgregor,poole2017chemical}.\par
Due to the associative feedback mechanism that catalyses the production of long term memory $L$, the system exhibits an automatic adaptation to the strength of the association between $S$ and $T$ (controlled by the environmental parameter $\epsilon$). The level of $L$ essentially becomes the learned estimation of the strength of this association. Hence $L$ is a discrete, low-pass filtered version of the time evolution of $\epsilon(t)$. The cut-off frequency of this filter is governed by the characteristic time scales of the network.\par
Tuning of our system parameters is based only on the temporal dynamics of the environment (period of delivery, spacing between stimulus and toxin, decay rate of the toxin, characteristic time of environmental changes). When all these factors are combined, only one parameter remains, and governs the trade-off between residual toxin level and the amount of antidote produced. This parameter can be tuned based on the relative effect of toxin damage and antidote cost.\par
Note that in addition to lacking a learning ability, the pre-emptive network requires tuning to the strength of the toxin, whereas the associative network is self-tuning. The associative network could be less resilient if the toxin delivery events were short and fast, but there would be no favourability for very slow decay of long term memory, so this only holds under extreme parameter values. In this work, the focus is situations beyond simple responses, where there is a tangible value in anticipation and learning. For scenarios where learning has no selective advantage, we would of course not expect any learning systems to emerge, so those scenarios are not relevant to the present work.\par
Beyond illustrating chemical associative learning, we presented simulations of emergent dissipative structures that can exploit these learning abilities to defend themselves against toxic compounds. These RDSTs exhibit a diverse repertoire of behaviours including replication (autocatalysis) \cite{pearson}, inter-structure competition \cite{bartlett2015emergence}, homeostasis and symbiosis \cite{bartlett2014life,bartlett2016precarious}, among others \cite{awazu2004relaxation,froese2,gagnon2015small,gagnon2018selection,lesmes2003noise,nishiura2001spatio,stich2013parametric}. With the results presented herein, we can now add associative learning to the remarkable list of life-like properties exhibited by RDSTs. In fact, they now satisfy the elementary four conditions to be classed as `lyfe' \cite{Bartlett_2020}.\par
The addition of a simple learning network allows these emergent patterns to exploit environmental correlations to encode information (a single variable in this case), and use that information to adapt their behaviour. \autoref{fig:UniformAssociativeToxin} and \autoref{fig:DiffusionAssociativeToxin} illustrate how toxin delivery initially degrades the spot population until long term memory builds up. This promotes the synthesis of antidote compound, and the spots then replicate into regions from which they had previously been eliminated.\par
In addition, the system shows a degree of functional differentiation, in which the spots that are exposed to toxin and stimulus build up a local memory and produce antidote, whereas isolated spots in the core of the group are largely shielded by the toxin (in the boundary delivery case, which is spatially heterogeneous in terms of the delivery of toxin and stimulus). This illustrates that specialisation in a colony of dissipative structures naturally follows from differences in environmental selective forces, and does not require genetically-encoded, cellular structures or complex molecular machinery. A similar spatial division of function was observed in thermal homeostasis models of RDSTs \cite{bartlett2016precarious}. In the context of prebiotic structure formation, one could even speculate that environmental spatial heterogeneity may be the driving force for the emergence of individuated patterns. Perhaps the differing physical selective forces placed upon such heterogeneous systems naturally promotes the emergence of boundary structures.\par
A natural question that arises from our results is the degree to which the emergent behaviours are genuinely emergent. One might contend that this system is simply a hand-designed reaction network that necessarily achieves the desired goal because it was engineered to do so. Thus how does this have any relevance to the natural world? The relevance of our results lies in the demonstration that such a simple chemical learning mechanism is possible and the conditions under which it is dynamically favoured. What we have presented is a locus from which the feasibility of emergent learning in OoL contexts can be explored. Having demonstrated the core requirements for autonomous chemical learning, the next question to answer is how such behaviour would still be favoured in the presence of larger networks populated by deleterious side reactions that are neutral or detrimental to learning? This would constitute a rigorous test of the emergence of the demonstrated learning behaviours.\par
Specifically, we intend to embed our system in progressively larger networks containing random reactions. The null hypothesis of such an investigation would be that adding superfluous reactions to learning networks disrupts and obscures the learning behaviour, and the system simply loses its learning abilities. Rejection of such a hypothesis would require the observation of conditions in which learning is dynamically favoured by a set of external conditions, i.e., a positive feedback between learning and the resilience or persistence of the system performing the learning. In this case there would be a subtle interplay between thermodynamic, kinetic and higher level effects that is difficult to predict without numerical or experimental modelling. The second law of thermodynamics is of limited utility because the system is not isolated and short term or stochastic effects can be amplified non-linearly. However, the system's dynamics on average would be exergonic. Hence our next phase of modelling will explore the selection of learning behaviour from a larger, naturally disordered space of dynamical processes.\par
We presented these results in the context of a proto-metabolic system that is vulnerable to degradation by a toxic compound. This compound's arrival is time varying but predictable via the occurrence of its precursor compound in associated environments. While appearing somewhat esoteric, this scenario could be generalised to other situations, such as predicting the arrival of nutrient compounds, or any exploitable association of time-varying events. If a learning system existed in an environment with no learnable features, then clearly a learning ability would be futile, offer no selective advantage, and would not persist (instantaneous feedback responses with no memory would be the optimal strategies in that case). However, life is surrounded by learnable features, and despite the stochastic and challenging nature of the Hadean and Archaen eons, there would still have been periodic, correlated, or predictable features of the environment that a nascent prebiotic system could potentially exploit (at the very least, diurnal or tidal cycles that would correlate with external fluxes of different compounds).\par
The primary feature separating life from the abiotic world is its unique information-processing abilities \cite{baluvska2016having,ben2009learning,ben2014physics,delgado1997collective,erez2017communication,farnsworth2013living,gershman2021reconsidering,ginsburg2010evolution,hopfield1994physics,kirchhoff2017there,manicka2019cognitive,marzen2018optimized,mitchell2009adaptive,sorek2013stochasticity,tkavcik2016information,watson2016evolutionary}, whereas non-living entities are at best only weakly predictive, teleological, nor do they exhibit perception-action cycles or measurement-feedback protocols \cite{boyd2017leveraging,brittain2019biochemical,ito2015maxwell}. When considering the emergence of the living state, it seems clear that information-processing and learning are a fundamental feature that must arise at some point. Given its importance, we suggest that that point was relatively early in the abiogenesis story, and we hope that the results presented here may provide a fruitful path to understanding that emergence in a general context.\par
In the OoL field, tremendous progress has been achieved in the areas of droplet systems \cite{cejkova2014dynamics,cejkova2017chemotaxis,cejkova2017droplets,hanczyc2003experimental,hanczyc2,hanczyc2011metabolism,holler2019droplet} and coacervates \cite{booth2019spatial,de2004complex,jia2014rapid,jia2019membraneless,li2014synthetic,qiao2017predatory,tang2014fatty,williams2014spontaneous}. In both cases, non-lipid molecules spontaneously produce individuated structures or boundaries. Such structures readily exhibit complex behaviour such as self-replication, chemotaxis, molecular concentration, or the enhancement of oligimerisation reactions. We suggest that the incorporation of simple learning circuits as presented in this work could add a novel and essential feature to these protocellular systems. In addition, it is likely that our learning circuits could be implemented using peptide or RNA-based systems. This could add a whole new dimension to contemporary origins experiments.\par

\section{Conclusions}
\label{sec:conc}
This work presents a simple catalytic reaction network capable of encoding correlations in environmental variables. In the context of emergent, prebiotic systems, such an associative learning ability could provide a defense mechanism against detrimental, but predictable perturbations imposed by an unstable environment. We demonstrated how associative chemical learning can be implemented using long and short term memory compounds. This allows the system to sense its environment over different time scales and encode the presence of correlated features. The synthesis of long term memory effectively implements an AND operation on the short term memory and the environmental variable being sensed. Such a mechanism allows the long term memory concentration to serve as a low-pass filter on the environmental variable (the degree to which a toxin is correlated with a precursor compound). Hence the learning ability is continuous, as opposed to discrete.\par
We showed that such a learning capacity can enhance the resilience of emergent chemical patterns (RDSTs). When endowed with learning networks, these structures are capable of autocatalysis, homeostasis and rudimentary learning. Implementations of this chemical learning system in OoL models such as droplets, coacervates or mineral surfaces could introduce a novel level of adaptive functionality that is an essential step in the journey from geophysics and geochemistry to biophysics, biochemistry and information processing.\par
Future work will carry out rigorous tests on the persistence of learning in the presence of deleterious side reactions. Establishing these necessary and sufficient conditions for emergent learning would have significant implications for research in Artificial life, OoL, systems chemistry, complexity science and synthetic biology.

\section*{Acknowledgments}
We gratefully acknowledge Artemy Kolchinsky of the University of Tokyo and Sante Fe Institute for his insightful and constructive review of a pre-publication version of this manuscript. This work was supported by the Caltech Division of Geological and Planetary Sciences Discovery Fund. We wish to thank the Caltech GPS `Astrobiothermoinfoevo' group for the numerous inspiring discussions that helped catalyse this work. Similarly, we thank the various members of the Earth-Life Science Institute and EON program members for creating such an inspiring and creative environment for intellectual exploration.


\providecommand{\noopsort}[1]{}\providecommand{\singleletter}[1]{#1}%

\section*{Supplementary Information}
\label{sec:supp_info}
\begin{figure}[ht]
    \centering
    \includegraphics[width=0.8\linewidth]{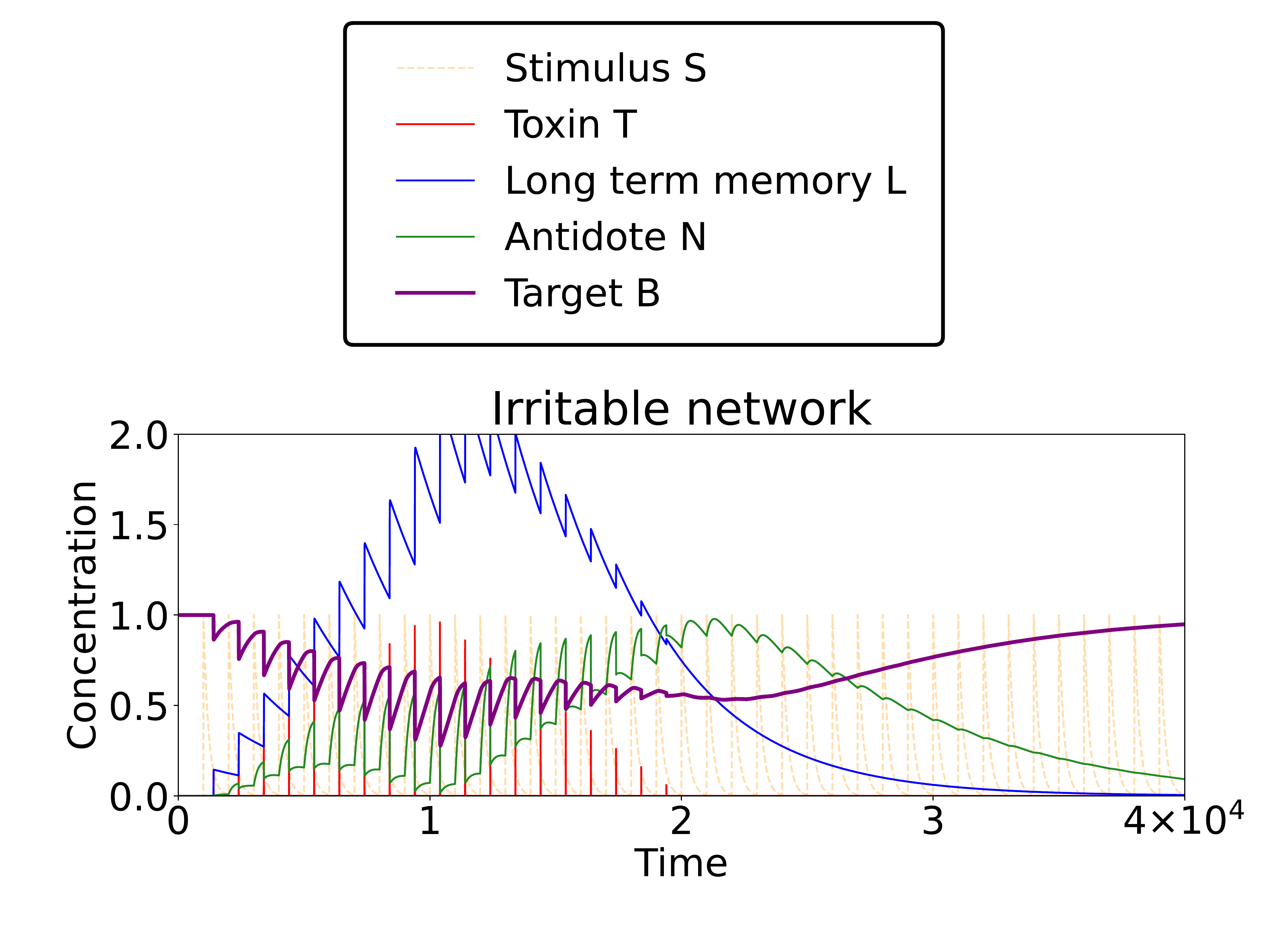}
    \caption{{\bf Dynamics of key compounds in lowly associative network $\mathcal{N}_{LL}$.}}
    \label{fig:L_T_nwk}
\end{figure}

In the following we consider a slightly streamlined version of our associative learning network that we refer to as the Lowly Associative Network $\mathcal{N}_{LL}$. This network omits the use of a short term memory, and simply produces long term memory in response to the presence of T via:
\begin{equation}
T \stackrel{k_{LL}}\longrightarrow L + T. \label{eq:assoL2}
\end{equation}
All other reactions of the associative network remain the same. The dynamics of the optimised version of this network are shown in \autoref{fig:L_T_nwk}.\par
We see that its performance is equivalent to the associative network, $\mathcal{N}_A$, due to the fact that long term memory need not track the association of S and T (though the overall network must be sensitive to S and T, otherwise it reduces to the direct network $\mathcal{N}_D$). It is necessary for the production of N to be pre-emptive and timely, such that the arrival of T can be mitigated. Hence the associative (using T to stimulate long term memory \textit{and} S to stimulate N) character of the lowly network is still essential to its resilience in this environment.\par
Given the above results, it is natural to question the necessity of the short term memory in the associative network $\mathcal{N}_A$. The environment used in this work was arguably the simplest possible environment that highlights a functional role for associative learning. However, in upcoming works we will be simulating environments that are more complex than this, involving more than one precursor and one toxin, and more complex time variabilities of the signals and their inter-relationships. In such environments, simply tracking the level of T is insufficient for an associative network to retain long term resilience. In fact, such networks must be more closely tied to the associations between environmental variables, and in order to be expandable, there must be the possibility for a broad range of associative memories to form.\par
It is for these reasons that we retain a short term memory in network $\mathcal{N}_A$, since future work will make use of elaborated versions of this elementary network motif.

\end{document}